\documentclass[aps,prd,reprint,superscriptaddress]{revtex4-1}
\usepackage{hyperref}
\usepackage{graphicx}
\usepackage{enumerate}
\hypersetup{
    pdfnewwindow=true,      % links in new window
    colorlinks=true,       % false: boxed links; true: colored links
    linkcolor=blue,          % color of internal links
    citecolor=black,        % color of links to bibliography
    filecolor=blue,      % color of file links
    urlcolor=blue           % color of external links
}

\usepackage{subfigure}
\usepackage{amsmath}
\usepackage{amssymb}
\usepackage{amsfonts}
\RequirePackage{lineno}
\usepackage{color}
\usepackage{enumitem}

\def\be{\begin{equation}}
\def\ee{\end{equation}}

\providecommand{\apj}[0]{Astrophys. J.}
\providecommand{\apjl}[0]{Astrophys. J. Lett.}

\providecommand{\mnras}[0]{Mon. Not. Roy. Astron. Soc. }

\providecommand{\prd}{Phys. Rev. D.}

\providecommand{\jcap}[0]{JCAP}

\begin{document}

\title{Can a Single High-energy Neutrino from Gamma-ray Bursts be a Discovery?}

\author{Imre Bartos}
\email[]{ibartos@phys.columbia.edu}
\affiliation{Department of Physics, Columbia University, New York, NY 10027, USA}
\author{Szabolcs M\'arka}
\affiliation{Department of Physics, Columbia University, New York, NY 10027, USA}

\begin{abstract}
Current emission models of GeV-PeV neutrinos from gamma-ray bursts (GRBs) predict a neutrino flux with $\ll 1$ detected neutrinos per GRB with kilometer-scale neutrino observatories. The detection of this flux will require the stacking of data from a large number of GRBs, leading to an increased background rate, decreasing the significance of a single neutrino detection. We show that utilizing the temporal correlation between the expected gamma-ray and neutrino fluxes, one can significantly improve the neutrino signal-to-noise ratio. We describe how this temporal correlation can be used. Using realistic GRB and atmospheric neutrino fluxes and incorporating temporal, spectral and directional information, we estimate the probability of a single detected GRB-neutrino being a 5$\sigma$ discovery.
\end{abstract}

\maketitle

\section{Introduction}

The emission mechanism that produces gamma-ray bursts (GRBs) is not yet well understood~\cite{2014arXiv1407.5671M}. High-energy (GeV-PeV) neutrinos may also be produced in GRBs, and their detection would help reveal the emission processes in relativistic GRB jets~\cite{1997PhRvL..78.2292W,2011PhRvL.106n1101A,2012Natur.484..351A,2013PhRvL.110x1101B}. Additionally, high-energy neutrino observations probe the contribution of GRBs to the observed cosmic ray flux~\cite{2012Natur.484..351A}, and are used in multimessenger searches with, e.g., gravitational waves \cite{2013JCAP...06..008A,2014arXiv1407.1042T}.

The search for neutrinos from GRBs is one of the primary goals of existing and planned high-energy neutrino observatories. The most sensitive, recent GRB-neutrino analyses, carried out using the IceCube detector~\cite{2006APh....26..155I}, already provide meaningful constraints on the neutrino emission~\cite{2011PhRvL.106n1101A,2012Natur.484..351A}. Nevertheless, there has been no confirmed detection of GRB-neutrinos so far.

Given the small predicted flux, the focus of GRB-neutrino searches has been to maximize the number of detectable neutrinos, by stacking data for a large number of GRBs and by considering an extended time window around GRBs in which neutrinos are looked for. In the latest IceCube analysis \cite{2012Natur.484..351A}, the search time window included the interval between the earliest and latest reported gamma-ray emission. Additionally, a \emph{model-independent} analysis was carried out for neutrinos within $\pm1$\,day around each GRB.

The most recent flux estimates for GRBs predict a relatively low number of detected neutrinos compared to earlier calculations~\cite{1997PhRvL..78.2292W,2004APh....20..429G}. Considering the standard fireball-internal shock model, H\"{u}mmer \emph{et al.}~\cite{2012PhRvL.108w1101H} obtains the expected TeV-PeV neutrino flux from observed GRB properties by numerically modeling the emission process, allowing for a detailed microphysical analysis. They find that constraining the GRB emission parameters will require an extended, multi-year observation with IceCube. An alternative GRB emission scenario with significant neutrino production is sub-photospheric dissipation of relativistic GRB jets by the proton-neutron collisions~\cite{2010MNRAS.407.1033B}. This model, which robustly reproduces the observed GRB photon spectrum, also predicts the emission of $\sim$\,GeV neutrinos~\cite{2000ApJ...541L...5M} that could be detectable by IceCube and its low-energy extension DeepCore~\cite{2012APh....35..615A} over a period of a few years~\cite{2013PhRvL.110x1101B}. With these GRB-neutrino rate estimates, the extended searches required for detection will also have a significant contribution from atmospheric background neutrinos. Therefore, it can be critical to use all available information about the emission process to better identify astrophysical neutrinos by improving the signal to noise ratio (see, e.g., \cite{2008APh....28..540V,2013APh....50...57B} for improved stacking methods).

Both gamma-rays and neutrinos are expected to be produced in relativistic outflows (jets) with typical Lorentz factor $\Gamma\gtrsim100$. Due to this relativistic expansion and their closely connected production mechanism, gamma-ray and high-energy neutrino fluxes will be temporally correlated in the observer frame. For instance, internal shocks in relativistic outflows can accelerate both protons and electrons. These protons lose some of their energy to pion production, emitting high-energy neutrinos, while the electrons can produce the observed gamma-ray emission due to synchrotron or inverse-Compton radiation \cite{1997PhRvL..78.2292W}. In this model, the emission of neutrinos and gamma-rays will be both synchronous with the internal shock in the observer frame. The temporal correlation also applies for cases in which the emission region is different, e.g., for photospheric gamma-ray emission, as long as the jet front is also relativistic. In this case, energy dissipation in the jet below the photosphere produces both neutrinos and an $e^{\pm}$ plasma, and the latter radiates its energy in gamma rays after reaching the photosphere. Since the dissipation region advances relativistically, there will be essentially no delay between the produced neutrinos and gamma-rays in the observer frame.

We note that mildly relativistic jets, or jets that are still beneath a stellar envelope, may produce high-energy neutrinos that precede the observable gamma-ray emission~(e.g., \cite{2003PhRvD..68h3001R,2012PhRvD..86h3007B,Baret20111,2013PhRvL.111l1102M}). Another exception is temporally extended $\sim$\,GeV gamma-ray emission, which can be connected to $\gg$\,PeV neutrino emission to which IceCube is less sensitive for sources towards the Northern hemisphere \cite{2012ApJ...757..115A}.  While these are also promising sources of neutrinos, in the following we focus on the $\sim$\,MeV gamma-ray emission from luminous GRBs for which there is no appreciable delay. This lack of delay enables the use temporal correlation without information on the source structure or its redshift.

In this paper, we examine whether even a single detected GRB-neutrino could be a discovery. Such a single neutrino would allow for the earliest possible detection, without the need to wait for the accumulation of multiple signal neutrinos. We make use of the temporal correlation expected between gamma-ray and neutrino emission in GRBs to show that even one astrophysical neutrino can be highly significant.

\begin{figure}
\begin{center}
\resizebox{0.49\textwidth}{!}{\includegraphics{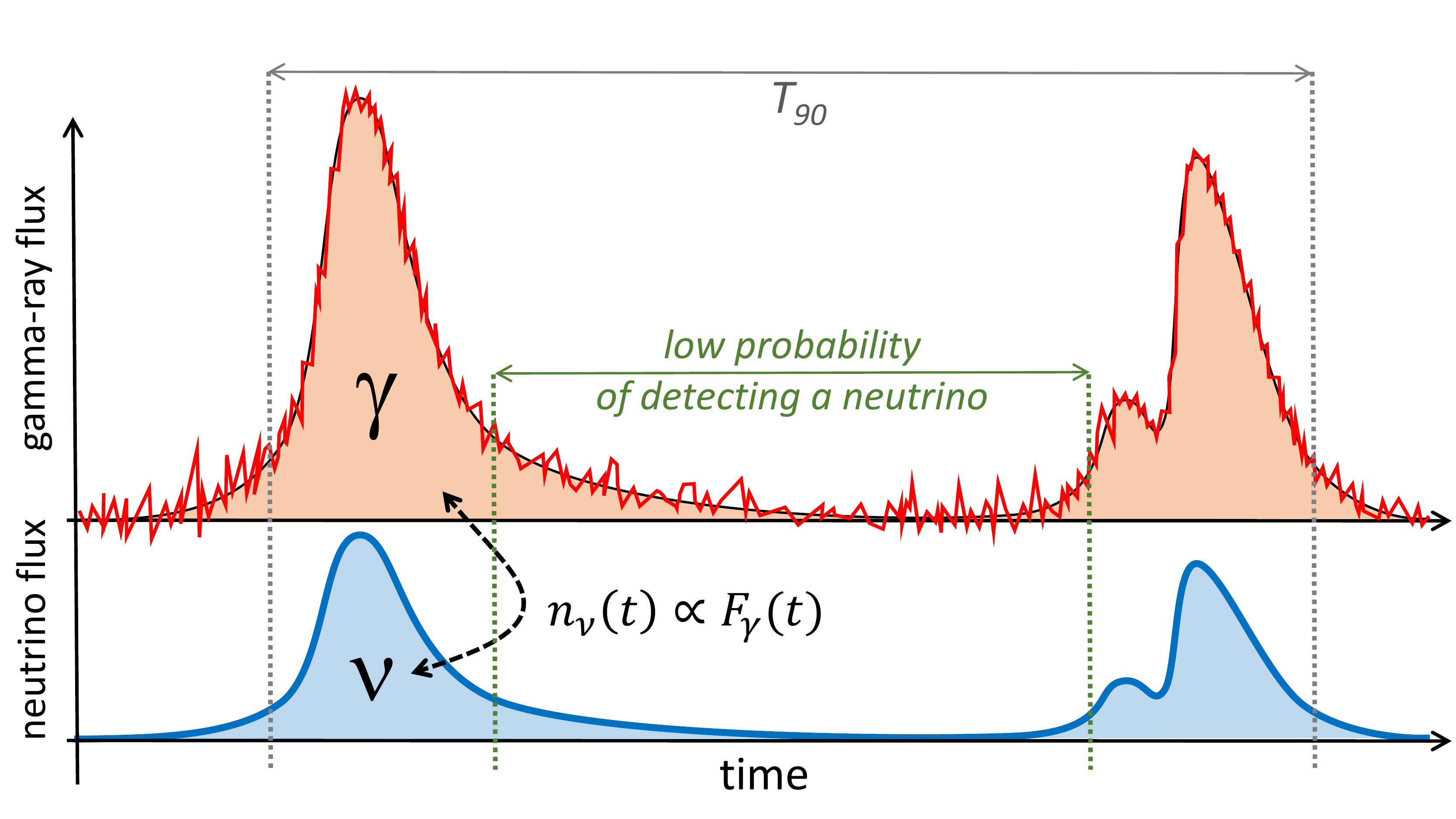}}
\end{center}
\caption{Schematic diagram of the observed gamma-ray flux (top) and the expected neutrino flux (bottom) from a GRB. The presented method assumes that the rate of detected neutrinos ($n_\nu$) is expected to be proportional to the gamma-ray flux ($F_\gamma$). Accounting for this proportionality can significantly decrease the false detection rate of background neutrinos.}
\label{figure:schematic}
\end{figure}

\section{Significance of gamma-ray-neutrino temporal correlation}
\label{section:tempcorr}

GRB neutrinos are more likely to be detected when the gamma-ray flux is higher. Below we utilize this correlation to associate an increased significance with astrophysical neutrinos by reducing the false alarm rate.

In general, the expected rate of detected neutrinos, $\dot{n}_{\nu}$, can depend on the gamma-ray flux $F_{\gamma}$ as well as other properties of the GRB, e.g., the photon power spectrum and redshift. In the following, we assume $\dot{n}_{\nu} \propto F_{\gamma}$.  To justify the use of this proportionality, we consider the neutrino emission estimates of H\"{u}mmer \emph{et al.}~\cite{2012PhRvL.108w1101H}, who calculated the neutrino fluence and spectrum for 117 GRBs that were used for a GRB-neutrino search with partially completed IceCube~(40-strings; \cite{2011PhRvL.106n1101A}), covering a 1-year observation period and GRBs in the Northern hemisphere.  To calculate the expected number $n_{\nu}$ of (muon) neutrinos detected by IceCube, we take the effective area of the full IceCube detector (for muon neutrinos arriving from a declination $>60^\circ$; \cite{2014arXiv1406.6757I}, which is more conservative than using a direction dependent effective area). We find that the obtained $n_{\nu}$ values for the 117 GRBs can indeed be characterized by the GRB fluences $S_{\gamma}$, following the proportionality
\begin{equation}
n_{\nu} \approx 10^{-2} \left(\frac{S_{\gamma}}{10^{-4}\,\mbox{erg\,cm}^{-2}}\right)
\label{equation:fluenceneutrino}
\end{equation}
with coefficient of determination $R^2=0.65$. This shows that $S_{\gamma}$ is a good predictor of $n_{\nu}$. For an actual GRB-neutrino search, one can adopt the calculated $n_{\nu}$ for each GRB in the analysis, which can further increase the advantage of assigning neutrino significance based on the measured GRB flux.

With $\dot{n}_{\nu} \propto F_{\gamma}$, a natural choice of the neutrino test statistic is $F_{\gamma}(t_{\nu}) / \dot{n}_{\rm bg}(\Omega_{\nu})$, where $F_{\gamma}(t_{\nu})$ is the observed GRB flux at the time $t_{\nu}$ of the detected neutrino, $\Omega_{\nu}$ is the reconstructed direction of the neutrino, and $\dot{n}_{\rm bg}$ is the direction dependent atmospheric background neutrino detection rate. In the following, for simplicity, we neglect the directional dependence of $\dot{n}_{\rm bg}$, and consider $F_{\gamma}(t_{\nu})$ as the neutrino test statistic (see, e.g., \cite{2011ApJ...732...18A} for the weak directional dependence of the background rate). Adopting this test statistic, one can calculate the p-value of an observed neutrino. Consider a set of GRBs that are the subject of the analysis. For each GRB, one can identify a time interval within which neutrinos are searched for, e.g., $T_{90}$, during which $90\%$ of the gamma-ray fluence is observed. Let $T$ be the total observation duration for all GRBs combined. The p-value of an observed neutrino that is temporally coincident with $F_{\gamma}(t_{\nu})$ gamma ray flux can then be defined as
\begin{equation}
p_{\nu} = \frac{1}{T}\int_{F_{\gamma}(t_{\nu})}^{\infty} \frac{\partial T(F_{\gamma}')}{\partial F_{\gamma}'}  dF_{\gamma}',
\label{equation:pvalue}
\end{equation}
where $(\partial T(F_{\gamma})/\partial F_{\gamma}) dF_{\gamma}$ is the duration during which the observed gamma-ray flux was within $[F_{\gamma}, F_{\gamma}+dF_{\gamma}]$ for all GRBs considered.

\section{Sensitivity}

The above test statistic can be used to filter out background neutrinos by requiring a low false alarm rate (FAR) for the search using a threshold flux $F_{\gamma}^{\rm min}$. To assess the sensitivity of the test statistic, here we quantify the false dismissal rate (FDR) as a function of $F_{\gamma}^{\rm min}$. First, let us consider the simple case in which $F_{\gamma}$ is precisely known for all GRBs, i.e. if background fluctuations were negligible. In this case, the FDR corresponding to $F_{\gamma}^{\rm min}$ is simply related to the total gamma-ray fluence integrated over fluxes below $F_{\gamma}^{\rm min}$:
\begin{equation}
\mbox{FDR}[F_{\gamma}^{\rm min}] = \frac{\mbox{FDR}_0}{S_{\rm total}}\int_{F_{0}}^{F_{\gamma}^{\rm min}} \frac{\partial T(F_{\gamma}')}{\partial F_{\gamma}'} F_{\gamma}'  dF_{\gamma}',
\label{FDR}
\end{equation}
where $\mbox{FDR}_0$ is the maximum rate of false dismissal (i.e. if all neutrinos were dismissed), $S_{\rm total}~=~\int_{F_{0}}^{\infty}\partial T(F_{\gamma}')/\partial F_{\gamma}' F_{\gamma}' dF_{\gamma}'$ is the total fluence of the GRBs in the analysis, and for the no-fluctuation case the minimum considered fluence can be chosen as $F_{0}=0$.

The presence of noise, however, complicates this estimate. Let the measured flux be $F_{\gamma} = F_{\rm grb} + \varepsilon_{\gamma}$, with $F_{\rm grb}$ being the real GRB flux and $\varepsilon_{\gamma}$ the random background flux drawn from a normal distribution with 0 mean and $\sigma_\gamma$ standard deviation. The background can be estimated from the observed flux before and after each GRB, therefore we assume that the mean of the noise is 0, and the standard deviation is known during the bursts.

Consider now the background-only case with $F_{\rm grb}=0$. The average measured flux in this case will be $\langle F_{\gamma}\rangle_{\rm bg} < 0$, since we exclude the highest values from the average above the threshold $F_{\gamma}^{\rm min}$. To ensure that the background fluence estimate will be 0, we can modify the sub-threshold fluence estimate in Eq.\,(\ref{FDR}) by taking $F_{0}=-F_{\gamma}^{\rm min}$. We can calculate this new fluence estimate as a function of the (unknown) GRB flux:
\begin{equation}
F_{\gamma}^{\rm estimate} = \frac{1}{P_0 \sqrt{2 \pi} \sigma_{\gamma}}\int_{-F_{\gamma}^{\rm min}}^{F_{\gamma}^{\rm min}} e^{-\frac{(F_{\gamma}' - F_{\rm grb})^2}{2\sigma_{\gamma}^{2}}}   F_{\gamma}'   dF_{\gamma}',
\label{equation:estimate}
\end{equation}
with
\begin{equation}
P_0=\frac{1}{\sqrt{2 \pi} \sigma_{\gamma}}\int_{-F_{\gamma}^{\rm min}}^{F_{\gamma}^{\rm min}} e^{-\frac{(F_{\gamma}' - F_{\rm grb})^2}{2\sigma_{\gamma}^{2}}} \, dF_{\gamma}'
\end{equation}, is accurate for the background, but it still underestimates the GRB flux; one can see that $F_{\gamma}^{\rm estimate}<F_{\rm grb}$.  This bias, nevertheless, is only significant for threshold values that are comparable to the background fluctuations, and our estimate will be correct for greater thresholds (see below).

\section{Results}

To determine the connection between FDR and FAR, we use the light curves of the 583 GRBs detected by Swift-BAT prior to December 2013 \footnote{\url{http://swift.gsfc.nasa.gov/archive/grb_table/}}.  We use their light curves during the time periods identified by the detector to be within the GRBs' $T_{90}$. We estimate the GRBs fluence by integrating their flux, which is an underestimate of their actual flux given the limited sensitive energy band of Swift-BAT (15\,keV\,$\leq$\,$E$\,$\leq$\,350\,keV). We bin the data within 5-s intervals (see below).  For simplicity, we use a uniform background uncertainty $\varepsilon_{\gamma}$ over the whole data set. A detailed GRB-neutrino search can use different uncertainties for each burst.

\begin{figure}
\begin{center}
\resizebox{0.49\textwidth}{!}{\includegraphics{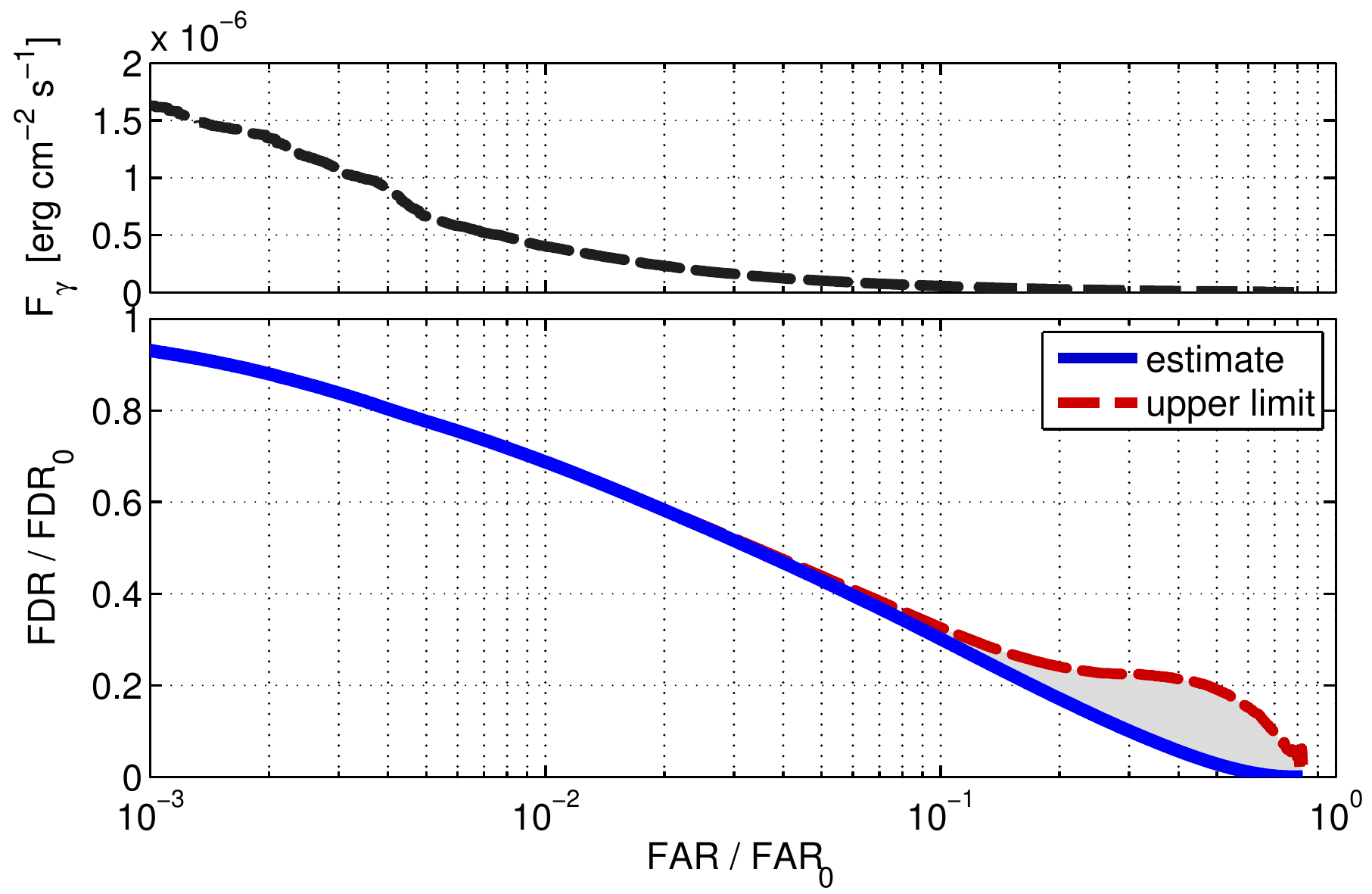}}
\end{center}
\caption{\emph{Lower panel:} False dismissal rate of high-energy neutrinos from GRBs due to requiring the contemporaneous gamma-ray flux to exceed a threshold, as a function of the false alarm rate corresponding to the same threshold (solid line).  FDR and FAR are both normalized by the total GRB and background neutrino rates, respectively. An FDR upper limit is also shown (dashed line) based on assuming a flat gamma-ray flux distribution as a function of time (see text).  \emph{Upper panel:} Gamma-ray flux threshold as a function of the corresponding false alarm rate. }
\label{figure:ROC}
\end{figure}

For a given FAR, it is straightforward to determine the corresponding gamma-flux threshold $F_{\gamma}^{\rm min}$ using Eq.\,(\ref{equation:pvalue}), since FAR$=$FAR$(F_{\gamma}^{\rm min})=p_{\nu}(F_{\gamma}^{\rm min})/T$. This threshold in turn is used to estimate the FDR using Eq.\,(\ref{FDR}). In Fig.\,\ref{figure:ROC} we show the obtained FDR as a function of FAR, both normalized to the total number of signal and background neutrinos, respectively, within the considered GRB time window. For comparison, we also show the dependence of $F_{\gamma}^{\rm min}$ on FAR.  One can see that, as expected, FAR decreases much faster than FDR, i.e. one can significantly improve the significance of detected astrophysical neutrinos with relatively little loss.

To determine the precision of the FDR estimate, we find a conservative upper limit on FDR by considering a uniform GRB flux distribution between $F_{\rm grb}=0$ and $F_{\rm grb}\gg\sigma_\gamma$. This uniform distribution gives an overestimate of the FDR compared to any realistic GRB flux distribution that is a decreasing function of the flux.  To see that this is indeed an upper limit, consider a data point with $F_{\rm grb}$ flux from the GRB.  The probability of this data point falling within $\pm F_{\gamma}^{\rm min}$ is $P_{0}$ (defined above).  The average contribution of this data point to the FDR is $\propto P_0 F_{\gamma}^{\rm estimate}$, and the discrepancy between the actual and identified falsely dismissed fluxes is $\approx P_0\,(F_{\rm grb} -  F_{\gamma}^{\rm estimate})$.  This contribution increases from $F_0=0$ up to $F_0 = 2\sigma_{\gamma}$, and then decreases afterwards.  Consequently, among GRB flux distributions that do not increase with $F_{\rm grb}$, the largest discrepancy is obtained for uniformly distributed fluxes.  Fig.\,\ref{figure:ROC} shows this upper limit for FDR by taking the measured $F_{\gamma}^{\rm estimate}$ and accounting for the potential underestimation assuming uniform flux distribution. As expected the discrepancy will be small for  $F_{\gamma}^{\rm min} \gtrsim 2\sigma_{\gamma} \approx 2\times10^{-8}$\,erg\,cm$^{-2}$\,s$^{-1}$ for FAR$/$FAR$_0$\,$\lesssim0.25$, making Eq.~(\ref{equation:estimate}) accurate for this range.

\section{Chance for discovery}

We now examine whether a single GRB-neutrino can be a 5$\sigma$ discovery if we use all available information to filter out the background. We implement a Monte Carlo simulation to obtain the rate and significance distribution of astrophysical neutrinos.  We use an unbinned likelihood ratio search that can be used for GRB-neutrino searches, incorporating temporal correlation beyond spectral and directional information. The steps of the analysis are the following:
\setlist[description]{font=\normalfont\itshape\space}
\begin{description}[leftmargin=\parindent,topsep=0pt,partopsep=3pt,parsep=0pt,itemsep=3pt,
    listparindent=\parindent]
\item[(i)] For a neutrino detected at time $t_\nu$, assign a temporal signal probability density function (PDF) that is proportional to the observed gamma-ray flux: $\mbox{P}^{\rm s}(t_{\nu}) \propto F_\gamma(t_\nu)$. The temporal background PDF is uniform.
\item[(ii)] Assign signal and background PDFs based on the neutrino energy and direction similar to GRB-neutrino analyses that use more general temporal emission models (e.g., \cite{2011PhRvL.106n1101A,2012Natur.484..351A}).
\item[(iii)] Assign a likelihood ratio $\mathcal{L}(t_\nu,\varepsilon_\nu, \overrightarrow{x}_\nu)$ to the neutrino in which the signal and background likelihoods are the products of three independent PDFs based on the neutrino's time of arrival $t_\nu$, reconstructed energy $\varepsilon_\nu$ and direction $\overrightarrow{x}_\nu$.
\item[(iv)] The significance of the neutrino, in comparison to a simulated background distribution, will be accurate for gamma-ray fluxes $F_\gamma \gtrsim 2 \sigma_\gamma$.
\end{description}
For the astrophysical neutrino energy distribution and detection rate from $p\gamma$ interactions in internal shocks, we adopt the energy distribution obtained by H\"{u}mmer \emph{et al.}~(Fig. 3 of \cite{2012PhRvL.108w1101H}, flux for all GRBs). For the energy distribution of detected neutrinos, we cross-correlate this distribution with the effective area of IceCube (see above). To obtain the rate of detected astrophysical neutrinos, we normalize the energy distribution using Eq.\,(\ref{equation:fluenceneutrino}). For background neutrinos, we adopt the energy spectrum by Abbasi \emph{et al.}~\cite{2011PhRvD..83a2001A}.  We estimate the rate of background neutrinos using the effective area of IceCube. We only consider background neutrinos with energy $\varepsilon_\nu\geq20$\,TeV, i.e. the lowest considered signal neutrino energy~\cite{2012PhRvL.108w1101H}.

To take into account the angular difference, we adopt a characteristic directional uncertainty of $\sigma_\nu=1^\circ$ for neutrinos, and $\sigma_{\rm grb}=0^\circ$ as well as $0^\circ$ for GRBs. We consider a normal distribution for the angular difference with zero mean and $(\sigma_\nu^2+\sigma_{\rm grb}^2)^{1/2}$ standard deviation (see, e.g., \cite{2011PhRvL.106n1101A}). We assume a uniform background directional distribution over the Northern hemisphere.

The PDF for the temporally coincident gamma-ray flux are considered as a function of FAR. The signal PDF is taken to be $\propto d\mbox{FDR}/d\mbox{FAR}$ from Fig.\,\ref{figure:ROC}, while the background PDF is a uniform function of FAR.

Assuming that the 117 GRBs considered in Ref.~\cite{2011PhRvL.106n1101A} is a representative sample of detected GRBs over a 1-year period over half the sky, using the duration (T90) of these GRBs, the expected total number of background neutrinos ($\varepsilon_\nu\geq20$\,TeV) that are temporally coincident with GRBs is $\sim 0.06$\,/\,year (for $\varepsilon_\nu\geq0.1$\,TeV it is $\sim 9$\,/\,year).  For the expected number of signal neutrinos we get $\sim 0.14$. With this background rate, we find that a single detected GRB-neutrino from $p\gamma$ interactions will be a $5\sigma$ discovery with $50\%$ ($30\%$) probability for $\sigma_{\rm grb}=0^\circ$ ($3^\circ$).

We also calculate the expected probability of discovery for neutrinos emitted by the $pn$ collisional heating of the jet \cite{2010MNRAS.407.1033B} using IceCube-DeepCore.  We expand the analysis of Bartos \emph{et al.}~\cite{2013PhRvL.110x1101B}. Here, we omit the use of the neutrino energy due to its strong dependence on the varying GRB Lorentz factor.  With an all-sky detected neutrino background rate of $\sim3\times10^{-4}$\,s$^{-1}$~\cite{2013PhRvL.110x1101B}, the expected number of temporally coincident background neutrinos with GRBs will be $\sim1$\,yr$^{-1}$. For the case of collisional heating, a single GRB-neutrino is unlikely ($<1\%$) to be a 5$\sigma$ discovery, while it can be a 3$\sigma$ \emph{evidence} with $\sim20\%$ probability. Discovery will require multiple detected neutrinos, with the discovery potential ($50\%$ chance to 5$\sigma$ discovery) being an average of $\sim5$ detected GRB-neutrinos.

\section{Conclusion}

We quantified the advantage of a GRB-neutrino search that takes into account a strong temporal correlation between the observed gamma-ray and high-energy neutrino fluxes in GRBs. We find that utilizing this correlation can significantly reduce the FAR of a GRB-neutrino search with a modest increase in FDR. A $10^{-1}$, $10^{-2}$ and $10^{-3}$ reduction of the FAR requires $30\%$, $70\%$ and $93\%$ FDR, respectively.  With the the most recent high-energy neutrino emission models predicting a relatively low rate of astrophysical neutrinos at which the background level becomes important, such a FAR reduction can be critical for claiming detection.

We provided a detailed prescription on how the gamma-neutrino temporal correlation can be utilized in a neutrino search for noisy GRB light curves. We estimated the fraction of detected GRB neutrinos that can be 5$\sigma$ discoveries even if only a single astrophysical neutrino is observed. For the case of $p\gamma$ interactions in collisionless shocks, we find up to $50\%$ probability of being a discovery. For the case of $\sim$ GeV neutrinos from sub-photospheric collisional $pn$ heating, the higher parameter uncertainties mean that multiple neutrinos will be required for discovery. We find that, on average, $\sim5$ neutrinos will be 5$\sigma$ discovery with $50\%$ probability.

We thank IceCube's GRB working group, Peter M\'esz\'aros, Zsuzsa Marka for their useful feedback; and Philipp Baerwald for the calculated neutrino spectra. We are thankful for the generous support of Columbia University in the City of New York and the National Science Foundation under cooperative agreement PHY-0847182.

\bibliographystyle{h-physrev}
%\bibliography{Refs}

\end{document}